\documentclass[pss,fleqn]{w-art}
\usepackage{times}
\usepackage{w-thm}
\usepackage{amssymb}

\usepackage[]{graphicx}
\begin{document}

\DOIsuffix{theDOIsuffix}

\Volume{XX}
\Issue{1}
\Month{01}
\Year{2003}

\pagespan{3}{}

\Receiveddate{2004}
\Reviseddate{2004}
\Accepteddate{2004}
\Dateposted{2004}

\keywords{Random Matrix Theory, Gaussian Unitary Ensemble,
Level Spacing Distribution, Series Expansion}
\subjclass[pacs]{05.45.-a,  05.45.Mt,  05.40.-a, 74.40.+k}



\title[Level-spacing distributions]{Level-spacing 
distributions of the Gaussian unitary random matrix ensemble}


\author[U.~Grimm]{Uwe Grimm\footnote{Corresponding
     author: e-mail: {\sf u.g.grimm@open.ac.uk}, Phone: +44\,1908\,659991,
     Fax: +44\,1908\,652140}} \address{Applied
Mathematics Department, The Open University, Walton Hall, Milton
Keynes MK7 6AA, UK}
\begin{abstract}
Level-spacing distributions of the Gaussian Unitary Ensemble (GUE) of
random matrix theory are expressed in terms of solutions of coupled
differential equations.  Series solutions up to order $50$ in the
level spacing are obtained, thus providing a very good description of
the small-spacing part of the level-spacing distribution, which can be
used to make comparisons with experimental or numerical data. The
level-spacing distributions can be obtained by solving
the system of differential equations numerically.
\end{abstract}
\maketitle                   






\section{Introduction}

Ever since the pioneering works of Wigner and Dyson
\cite{Wigner,Dyson1,Dyson2,Dyson3,Dyson4,DM63}, random matrix theory
has been a major tool in the investigation of complex systems in
physics, see {\it e.g.}\/ Refs.~\cite{Haake,Guhr}. In particular, the
three ``classical'' random matrix ensembles, the Gaussian Orthogonal
Ensemble (GOE), the Gaussian Unitary Ensemble (GUE) and the Gaussian
Sympletic Ensemble (GSE) have been studied in much detail \cite{Mehta}
and have proven of relevance in various applications, not alone in
physics, see {\it e.g.}\/ \cite{TW} for examples.

One of the characteristic features that is frequently used in
investigations is the distribution of differences between eigenvalues.
These level-spacing distributions are universal quantities, in the
sense that they correctly describe the level-spacing distributions in
many complex systems. In many applications, experimentally observed or
numerically calculated discrete data, such as energy levels of large
nuclei, are analyzed and their spacing distribution is compared to
that of the appropriate random matrix ensemble, where the choice of
the appropriate ensemble is dictated by the symmetry of the system.
Besides the spacing between neighboring levels, also the spacing
between levels separated by a fixed number of levels, {\it i.e.},
next-nearest neighbors, next-next-nearest neighbors, {\it etc.}, has
been considered. A recent example of an astonishingly precise
agreement between energy levels of a Hamiltonian and random matrix
distributions was found in a simple tight-binding model of an electron
moving on a planar quasiperiodic graph \cite{ZGRS,SGRZ1,SGRZ2,GRSZ},
which is used to model electrons in quasicrystals \cite{GS1,GS2}. But
the distribution functions of random matrix theory are even of
relevance in pure mathematics; the paradigm of an application in
number theory is the distribution of zeros of Riemann's zeta function
on the critical line \cite{Odlyzko}.

However, there exists no simple closed form for the level-spacing
distributions of the above-mentioned ensembles of random matrix
theory. Thus, in many cases, the empirical data are actually compared
to a so-called Wigner surmise, which corresponds to the spacing
distribution of neighboring levels obtained from two-by-two random
matrices, rather than from the limiting distribution for matrices of
infinite size. For example, in the GOE case, the difference between
the Wigner surmise and the true GOE spacing distribution is rather
small, but the example of the quasiperiodic tight-binding model
\cite{ZGRS,SGRZ1,SGRZ2,GRSZ} shows that it may indeed be detected in
physical applications. For the GUE case, the situation is somewhat
better as the complete small spacing expansion for the nearest
neighbor spacing distribution is known \cite{Haake}. Numerical values,
series expansions and a Pad\'{e} analysis for the nearest-neighbor
spacing distributions can be found in
\cite{Mehta,Haake,DiHa}. However, level spacings beyond the
nearest-neighbor distributions, the spacing distributions of levels
separated by a number of other levels, have scarcely been
considered. However, explicit power series expressions are available
\cite{Mehta92}, although the coefficients in the series still involve
determinants that need to be evaluated.

Here we employ the approach of Ref.~\cite{BTW}, which is based on the
relation to differential equations of Painlev\'{e} type discovered by
Jimbo {\it at al.} \cite{JMMS}, to obtain series expansions or
numerical estimates of these functions, compare also
\cite{WFC,FW2,FW3,F} for connections between random matrices and
Painlev\'{e} transcendents.  In this paper, we generalize this
approach to calculate series expansions for small spacings up to order
$50$ for the higher level-spacing distributions of the GUE. The
expansions provide a detailed account of the small-spacing behavior of
these distributions, and thus can be applied to analyze data. The
exact coefficients involve ratios of large integer numbers, which
makes them somewhat difficult to present in a paper. Therefore, the
analytic expressions of the expansion coefficients are not included;
they are provided in form of {\sc Mathematica} \cite{Wolfram} code on
the author's homepage \cite{code}.  The other two classical ensembles,
the GOE and the GSE, can be treated in a similar fashion, although the
calculation is somewhat more involved \cite{BTW}. The results for
these two universality classes will be published separately.

Why are we interested in calculating these expansion to such high
order in $s$? There are several reasons. First of all, the data
suggest that the series have infinite radius of convergence, thus the
level-spacing distributions appear to be analytic functions of
$s$. This means that, the more coefficients we know, the larger the
interval where the truncated series gives a good approximation of the
spacing functions. Secondly, the higher level spacing distributions
have leading terms of increasing power in $s$, so in order to at least
obtain the leading terms of the higher spacing function we have to
expand all function during the calculation up to that given order in
$s$. In our case, expanding up to order $s^{50}$ just suffices to
gives us the leading terms of the spacing distribution of levels
separated by five other levels. Thirdly, although it is, in principle,
possible to integrate the system of coupled differential equations
numerically and to obtain ``numerically exact'' results for the
spacing functions, this does require the knowledge of at least the
leading terms in the expansions as we cannot integrate directly from
$s=0$. Finally, it turns out to be very difficult to obtain a
numerically exact distribution function in this way, because the
distribution functions arise as differences between functions that
grow exponentially with $s$, and thus requires high precision
numerics. In fact, the author was not able to integrate numerically
the set of equations for the spacing functions with four or five
intermediate levels, even though the algebraic computer package {\sc
Mathematica} \cite{Wolfram} in principle allows for arbitrary
precision arithmetics. 

The paper is organized as follows. After this introduction, the method
used to calculate the spacing distribution is briefly summarized,
following closely the discussion of Ref.~\cite{BTW}. After that,
the series expansion data are presented and their applicability is
discussed. We end with some concluding remarks.

\section{Level-spacing distributions}

The basic objects that usually are considered are the probabilities
$E_{n}(s)$ that an interval of length $s$ contains exactly $n$
eigenvalues. Here, $s$ denotes the energy-level spacing in units of
the mean level spacing. The gap probability $E_{0}(s)$ that an interval
of length $s$ contains no eigenvalue at all can be expressed as a
Fredholm determinant of a certain integral operator $K$ with a sine
kernel, see \cite{TW} and references therein. Jimbo {\it et al.}\/
\cite{JMMS} showed that this Fredholm determinant can be expressed as
\begin{equation}
D(s;\lambda) \; := \; \det\left( I - \lambda K\right) \; = \; 
\exp\left( \int\limits_{0}^{\pi s}\! dx\; \frac{\sigma(x;\lambda)}{x}
\right)
\label{eq:d}
\end{equation}
where the function $\sigma(x;\lambda)$, considered as a function of
the variable $x$ whereas $\lambda$ is regarded as a parameter, is a
solution of a Painlev\'{e} V differential equation
\begin{equation}
\Big(x\:\frac{d^{2}\!\sigma}{d x^{2}}\Big)^{2} \: + \; 
4\Big( x\:\frac{d\sigma^{}}{d x} - 
\sigma \rule{0pt}{20pt}\Big)
\Big[ x\:\frac{d\sigma}{d x} -\sigma + 
\Big( \frac{d\sigma}{d x}\Big)^{\! 2}
\Big] \;  = \;  0. 
\label{eq:pv}
\end{equation}
The parameter $\lambda$ is introduced via the boundary condition
\begin{equation}
\sigma(x;\lambda) \; = \; -\frac{\lambda}{\pi}\: x \; + \; o(x)
\label{eq:ic}
\end{equation}
for $x\rightarrow 0$. For the GUE, $D(s;\lambda)$ is the generating
function of the probabilities $E_{n}(s)$, thus
\begin{equation}
E_{n}(s) =  D_{n}(s),
\label{eq:en}
\end{equation}
where we defined
\begin{eqnarray}
D_{0}(s) &=&  D(s;1) \label{eq:d0}\\ 
D_{n}(s) &=&  
\left.\frac{(-1)^n}{n!}
\frac{\partial^n D(s;\lambda)}{\partial\lambda^n}\right|_{\lambda=1},
\quad n>0.
\label{eq:dn}
\end{eqnarray}
The level-spacing distributions $P_{n}(s)$, which are the
probability distributions to find two energy levels at a distance $s$
with $n$ levels lying in-between, are then obtained as
\begin{equation}
P_{n}(s) \; = \; \frac{d^2}{ds^2} \; \sum_{m=0}^{n}\: (n-m+1)\: E_{m}(s)
\label{eq:pn}
\end{equation}
where $P_{0}(s)$ is the usual nearest-neighbour spacing distribution.

Now, it is rather straightforward to derive series expansions for the
distributions $P_{n}(s)$ by making a power-series ansatz for the
solution $\sigma(x;\lambda)$ of the differential equation
(\ref{eq:pv}) with the appropriate boundary condition (\ref{eq:ic}).
The distributions $P_{n}(s)$ were computed with the commercial
algebraic computer program {\sc Mathematica} \cite{Wolfram}. Whereas
it is rather easy to calculate the first few series coefficients in
this way, a straightforward implementation of the equations in an
algebraic programming language like {\sc Mathematica} will generally
not succeed to calculate higher orders. In order to achieve this, the
calculation was split into smaller tasks and an iterative scheme was
used, calculating the expansions coefficients one by one. In this way,
it was possible to obtain the series expansions for the spacing
functions $P_{n}(s)$ up to order $s^{50}$. 

\section{Series expansions}

Small-$s$ expansions 
\begin{equation}
P_{n}(s) = \sum_{k=0}^{\infty}p_{n;k}s^k
\label{eq:ps}
\end{equation}
of the level-spacing distributions $P_{n}(s)$ (\ref{eq:pn}) can be
obtained as follows. As a first step, we define functions
$\sigma_{n}(x)$ by
\begin{eqnarray}
\sigma_{0}(x) &=& \sigma(x;1),\\
\sigma_{n}(x) &=& \left.
\frac{\partial^n \sigma(x;\lambda)}{\partial\lambda^n}\right|_{\lambda=1}.
\end{eqnarray}
The function $\sigma_{0}(x)$ satisfies the differential equation 
(\ref{eq:pv}) and behaves as
\begin{equation}
\sigma_{0}(x) \; = \; -\frac{1}{\pi}\: x \; + \; o(x)
\label{eq:ic0}
\end{equation}
for small $x$. This follows from Eq.~(\ref{eq:ic}), which also determines
the small-$x$ behavior of $\sigma_{n}(x)$ for $n>0$,
\begin{eqnarray}
\sigma_{1}(x) & = & -\frac{1}{\pi}\: x \; + \; o(x),
\label{eq:ic1}\\
\sigma_{n}(x) & = & o(x)\qquad n>1.
\label{eq:icn}
\end{eqnarray}
For any $N$, the functions $\sigma_{n}(x)$, $0\le n\le N$ satisfy a
coupled set of differential equations that are obtained by taking the
derivative of the original Painlev\'{e} equation (\ref{eq:pv}) with
respect to $\lambda$ and putting $\lambda=1$. For instance, for $N=2$
the set of differential equations becomes
\begin{eqnarray}
0 & = & (x \sigma''_{0})^{2} + 
4(x\sigma'_{0} - \sigma_{0})(x\sigma'_{0} -\sigma_{0} +{\sigma'_{0}}^{2}), 
\label{eq:pv0}\\
0 & = & x^2\sigma''_{1}\sigma''_{0} +
2(x\sigma'_{1}-\sigma_{1})(x\sigma'_{0}-\sigma_{0}+{\sigma'_{0}}^{2})
\nonumber\\
& & + 
2(x\sigma'_{0}-\sigma_{0})(x\sigma'_{1}-\sigma_{1}+2\sigma'_{1}\sigma'_{0}),
\label{eq:pv1}\\
0 & = & x^2\sigma''_{2}\sigma''_{0} + (x \sigma''_{1})^{2}\nonumber\\
& & + 
2(x\sigma'_{2}-\sigma_{2})(x\sigma'_{0}-\sigma_{0}+{\sigma'_{0}}^{2})
\nonumber\\
& & + 
4(x\sigma'_{1}-\sigma_{1})(x\sigma'_{1}-\sigma_{1}+2\sigma'_{1}\sigma'_{0})
\nonumber\\
& & + 
2(x\sigma'_{0}-\sigma_{0})(x\sigma'_{2}-\sigma_{2}+2\sigma'_{2}\sigma'_{0}
+2{\sigma'_{1}}^{2}).
\label{eq:pv2}
\end{eqnarray}
In order to obtain series expansions for the level-spacing
distributions $P_{n}(s)$ (\ref{eq:pn}) with $0\le n\le N$, we first
compute series expansions for the functions $\sigma_{n}(x)$ for
$0\le n\le N$. This is done by inserting polynomials
\begin{equation}
\sigma_{n}(x) = \sum_{k=0}^{K}c_{n;k} \left(\frac{x}{\pi}\right)^{k}
\end{equation}
of order $K$ into the set of differential equations obtained from
(\ref{eq:pv}), and implementing the initial conditions
(\ref{eq:ic0})--(\ref{eq:icn}) by setting $c_{n;0}=0$ for $0\le n\le
N$, $c_{0;1}=c_{1;1}=-1$ and $c_{n;1}=0$ for $2\le n\le N$. The
initial conditions (\ref{eq:ic0}) and (\ref{eq:ic1}) motivate the
convenient choice of the expansion variable $x/\pi$ rather than $x$.

{}From the expansions of the functions $\sigma_{n}(x)$, $0\le n\le N$,
we can derive expansions for $D_{n}(s)$ with $0\le n\le N$ by using
Eqs.~(\ref{eq:d}), (\ref{eq:d0}), and (\ref{eq:dn}). For example,
we have
\begin{eqnarray}
D_{0}(s) & = &
\exp\left( 
\int\limits_{0}^{\pi s}\! \frac{\sigma_{0}(x)\,dx}{x}\right)
\label{eq:d0exp}\\
D_{1}(s) & = & 
D_{0}(s)\:\int\limits_{0}^{\pi s}\! \frac{\sigma_{1}(x)\, dx}{x},
\\
D_{2}(s) & = &
D_{0}(s)\:\left[
\int\limits_{0}^{\pi s}\! \frac{\sigma_{2}(x)\, dx}{x} +
{\left(\int\limits_{0}^{\pi s}\! \frac{\sigma_{1}(x)\, dx}{x}\right)}^{2}
\right],
\end{eqnarray}
\clearpage
\begin{table}[htb]
\caption{Numerical values of expansion coefficients (\ref{eq:ps}) of
the level-spacing distributions $P_{0}$, $P_{1}$, $P_{2}$, $P_{3}$,
$P_{4}$, and $P_{5}$.\label{tab1}}
\begin{small}
\begin{tabular}{@{}r@{\quad}r@{}l@{\quad}r@{}l@{\quad}r@{}l@{\quad}r@{}l@{\quad}r@{}l@{\quad}r@{}l@{}}
\hline
\multicolumn{1}{c}{$k$}&
\multicolumn{2}{c}{$p_{0;k}$}&
\multicolumn{2}{c}{$p_{1;k}$}&
\multicolumn{2}{c}{$p_{2;k}$}&
\multicolumn{2}{c}{$p_{3;k}$}&
\multicolumn{2}{c}{$p_{4;k}$}&
\multicolumn{2}{c}{$p_{5;k}$}\\
\hline
 $0$ &  $0$& &
        $0$& &
        $0$& &
        $0$& &
        $0$& &
        $0$& \\
 $1$ &  $0$& &
        $0$& &
        $0$& &
        $0$& &
        $0$& &
        $0$& \\
 $2$ &  $3$&$.290$ &
        $0$& &
        $0$& &
        $0$& &
        $0$& & 
        $0$& \\
 $3$ &  $0$& &
        $0$& &
        $0$& &
        $0$& &
        $0$& &
        $0$& \\
 $4$ & $-4$&$.329$ &
        $0$& &
        $0$& &
        $0$& &
        $0$& & 
        $0$& \\
 $5$ &  $0$& &
        $0$& &
        $0$& &
        $0$& &
        $0$& & 
        $0$& \\
 $6$ &  $3$&$.052$ &
        $0$& &
        $0$& &
        $0$& &
        $0$& & 
        $0$& \\
 $7$ & $-2$&$.374\!\times\! 10^{-1}$ &
        $2$&$.374\!\times\! 10^{-1}$ &
        $0$& &
        $0$& & 
        $0$& &
        $0$& \\
 $8$ & $-1$&$.339$ &
        $0$& &
        $0$& &
        $0$& &
        $0$& & 
        $0$& \\
 $9$ &  $2$&$.104\!\times\! 10^{-1}$ &
       $-2$&$.104\!\times\! 10^{-1}$ &
        $0$& &
        $0$& &
        $0$& &
        $0$& \\
$10$ &  $4$&$.004\!\times\! 10^{-1}$ &
        $0$& &
        $0$& &
        $0$& & 
        $0$& &
        $0$& \\
$11$ & $-9$&$.088\!\times\! 10^{-2}$ &
        $9$&$.088\!\times\! 10^{-2}$ &
        $0$& &
        $0$& & 
        $0$& &
        $0$& \\
$12$ & $-8$&$.685\!\times\! 10^{-2}$ &
        $0$& &
        $0$& &
        $0$& & 
        $0$& &
        $0$& \\
$13$ &  $2$&$.491\!\times\! 10^{-2}$ &
       $-2$&$.491\!\times\! 10^{-2}$ &
        $0$& &
        $0$& &
        $0$& & 
        $0$& \\
$14$ &  $1$&$.446\!\times\! 10^{-2}$ &
       $-3$&$.450\!\times\! 10^{-4}$ &
        $1$&$.725\!\times\! 10^{-4}$ &
        $0$& & 
        $0$& &
        $0$& \\
$15$ & $-4$&$.827\!\times\! 10^{-3}$ &
        $4$&$.827\!\times\! 10^{-3}$ &
        $0$& & 
        $0$& & 
        $0$& &
        $0$& \\
$16$ & $-1$&$.974\!\times\! 10^{-3}$ &
       $ 2$&$.625\!\times\! 10^{-4}$ &
       $-1$&$.313\!\times\! 10^{-4}$ &
        $0$& & 
        $0$& &
        $0$& \\
$17$ &  $7$&$.038\!\times\! 10^{-4}$ &
       $-7$&$.038\!\times\! 10^{-4}$ & 
        $0$& &
        $0$& &
        $0$& &
        $0$& \\
$18$ &  $2$&$.411\!\times\! 10^{-4}$ &
       $-9$&$.932\!\times\! 10^{-5}$ &
        $4$&$.966\!\times\! 10^{-5}$ &
        $0$& &
        $0$& &
        $0$& \\
$19$ & $-8$&$.042\!\times\! 10^{-5}$ &
        $8$&$.042\!\times\! 10^{-5}$ &
        $0$& & 
        $0$& &
        $0$& &
        $0$& \\
$20$ & $-2$&$.884\!\times\! 10^{-5}$ &
        $2$&$.496\!\times\! 10^{-5}$ &
       $-1$&$.248\!\times\! 10^{-5}$ &
        $0$& & 
        $0$& &
        $0$& \\
$21$ &  $7$&$.418\!\times\! 10^{-6}$ &
       $-7$&$.418\!\times\! 10^{-6}$ &
        $0$& &
        $0$& & 
        $0$& &
        $0$& \\
$22$ &  $3$&$.527\!\times\! 10^{-6}$ &
       $-4$&$.714\!\times\! 10^{-6}$ &
        $2$&$.357\!\times\! 10^{-6}$ &
        $0$& & 
        $0$& &
        $0$& \\
$23$ & $-5$&$.659\!\times\! 10^{-7}$ &
        $5$&$.678\!\times\! 10^{-7}$ &
       $-2$&$.805\!\times\! 10^{-9}$ &
        $9$&$.351\!\times\! 10^{-10}$&
        $0$& &
        $0$& \\
$24$ & $-4$&$.301\!\times\! 10^{-7}$ &
        $7$&$.180\!\times\! 10^{-7}$ &
       $-3$&$.590\!\times\! 10^{-7}$ &
        $0$& &
        $0$& &
        $0$& \\
$25$ &  $3$&$.684\!\times\! 10^{-8}$ &
       $-3$&$.816\!\times\! 10^{-8}$ &
        $1$&$.983\!\times\! 10^{-9}$ &
       $-6$&$.610\!\times\! 10^{-10}$&
        $0$& & 
        $0$& \\
$26$ &  $4$&$.991\!\times\! 10^{-8}$ &
       $-9$&$.240\!\times\! 10^{-8}$ &  
        $4$&$.620\!\times\! 10^{-8}$ &
        $0$& &
        $0$& &     
        $0$& \\
$27$ & $-2$&$.210\!\times\! 10^{-9}$ &
        $2$&$.677\!\times\! 10^{-9}$ &
       $-7$&$.013\!\times\! 10^{-10}$ &
        $2$&$.338\!\times\! 10^{-10}$ &
        $0$& & 
        $0$& \\
$28$ & $-5$&$.357\!\times\! 10^{-9}$ & 
        $1$&$.038\!\times\! 10^{-8}$ &
       $-5$&$.189\!\times\! 10^{-9}$ &
        $0$& & 
        $0$& &
        $0$& \\
$29$ &  $1$&$.484\!\times\! 10^{-10}$ &
       $-2$&$.587\!\times\! 10^{-10}$ &
        $1$&$.655\!\times\! 10^{-10}$ &
       $-5$&$.517\!\times\! 10^{-11}$ &
        $0$& &
        $0$& \\
$30$ &  $5$&$.265\!\times\! 10^{-10}$ &
       $-1$&$.040\!\times\! 10^{-9}$ &
        $5$&$.198\!\times\! 10^{-10}$ &
        $0$& &
        $0$& &
        $0$& \\
$31$ & $-1$&$.363\!\times\! 10^{-11}$ &
        $3$&$.320\!\times\! 10^{-11}$ &
       $-2$&$.936\!\times\! 10^{-11}$ &
        $9$&$.788\!\times\! 10^{-12}$ &
        $0$& &
        $0$& \\
$32$ & $-4$&$.738\!\times\! 10^{-11}$ &
        $9$&$.430\!\times\! 10^{-11}$ &
       $-4$&$.715\!\times\! 10^{-11}$ &
        $0$& &
        $0$& &
        $0$& \\
$33$ &  $1$&$.532\!\times\! 10^{-12}$ & 
       $-4$&$.317\!\times\! 10^{-12}$ &
        $4$&$.178\!\times\! 10^{-12}$ &
       $-1$&$.393\!\times\! 10^{-12}$ &
        $0$& &
        $0$& \\
$34$ &  $3$&$.915\!\times\! 10^{-12}$ &
       $-7$&$.816\!\times\! 10^{-12}$ &
        $3$&$.908\!\times\! 10^{-12}$ &
       $-1$&$.072\!\times\! 10^{-16}$ &
        $2$&$.680\!\times\! 10^{-17}$ &
        $0$& \\
$35$ & $-1$&$.700\!\times\! 10^{-13}$ &
        $5$&$.011\!\times\! 10^{-13}$ &
       $-4$&$.966\!\times\! 10^{-13}$ &
        $1$&$.655\!\times\! 10^{-13}$ &
        $0$& &
        $0$& \\
$36$ & $-2$&$.980\!\times\! 10^{-13}$ &
        $5$&$.956\!\times\! 10^{-13}$ &
       $-2$&$.979\!\times\! 10^{-13}$ &
        $7$&$.275\!\times\! 10^{-17}$ &
       $-1$&$.819\!\times\! 10^{-17}$ &
        $0$& \\
$37$ &  $1$&$.703\!\times\! 10^{-14}$ &
       $-5$&$.084\!\times\! 10^{-14}$ &
        $5$&$.071\!\times\! 10^{-14}$ &
       $-1$&$.690\!\times\! 10^{-14}$ &
        $0$& &
        $0$& \\
$38$ &  $2$&$.095\!\times\! 10^{-14}$ &
       $-4$&$.191\!\times\! 10^{-14}$ &
        $2$&$.098\!\times\! 10^{-14}$ &
       $-2$&$.473\!\times\! 10^{-17}$ &
        $6$&$.184\!\times\! 10^{-18}$ &
        $0$& \\
$39$ & $-1$&$.516\!\times\! 10^{-15}$ &
        $4$&$.542\!\times\! 10^{-15}$ &
       $-4$&$.539\!\times\! 10^{-15}$ &
        $1$&$.513\!\times\! 10^{-15}$ &
        $0$& &
        $0$& \\
$40$ & $-1$&$.365\!\times\! 10^{-15}$ &
        $2$&$.732\!\times\! 10^{-15}$ &
       $-1$&$.372\!\times\! 10^{-15}$ &
        $5$&$.620\!\times\! 10^{-18}$ &
       $-1$&$.405\!\times\! 10^{-18}$ &
        $0$& \\
$41$ &  $1$&$.205\!\times\! 10^{-16}$ &
       $-3$&$.613\!\times\! 10^{-16}$ &
        $3$&$.612\!\times\! 10^{-16}$ &
       $-1$&$.204\!\times\! 10^{-16}$ &
        $0$& &
        $0$& \\
$42$ &  $8$&$.260\!\times\! 10^{-17}$ &
       $-1$&$.657\!\times\! 10^{-16}$ &
        $8$&$.379\!\times\! 10^{-17}$ &
       $-9$&$.603\!\times\! 10^{-19}$ &
        $2$&$.401\!\times\! 10^{-19}$ &
        $0$& \\
$43$ & $-8$&$.622\!\times\! 10^{-18}$ &
        $2$&$.586\!\times\! 10^{-17}$ &
       $-2$&$.586\!\times\! 10^{-17}$ &
        $8$&$.620\!\times\! 10^{-18}$ &
        $0$& &
        $0$& \\
$44$ & $-4$&$.659\!\times\! 10^{-18}$ &
        $9$&$.384\!\times\! 10^{-18}$ &
       $-4$&$.824\!\times\! 10^{-18}$ &
        $1$&$.316\!\times\! 10^{-19}$ &
       $-3$&$.291\!\times\! 10^{-20}$ &
        $0$& \\
$45$ &  $5$&$.600\!\times\! 10^{-19}$ &
       $-1$&$.680\!\times\! 10^{-18}$ &
        $1$&$.680\!\times\! 10^{-18}$ &
       $-5$&$.600\!\times\! 10^{-19}$ &
        $0$& &
        $0$& \\
$46$ &  $2$&$.459\!\times\! 10^{-19}$ &
       $-4$&$.994\!\times\! 10^{-19}$ &
        $2$&$.648\!\times\! 10^{-19}$ &
       $-1$&$.509\!\times\! 10^{-20}$ &
        $3$&$.772\!\times\! 10^{-21}$ &
        $0$& \\
$47$ & $-3$&$.324\!\times\! 10^{-20}$ &
        $9$&$.972\!\times\! 10^{-20}$ &
       $-9$&$.972\!\times\! 10^{-20}$ &
        $3$&$.324\!\times\! 10^{-20}$ &
       $-1$&$.483\!\times\! 10^{-26}$ &
        $2$&$.966\!\times\! 10^{-27}$ \\
$48$ & $-1$&$.221\!\times\! 10^{-20}$ &
        $2$&$.516\!\times\! 10^{-20}$ &
       $-1$&$.407\!\times\! 10^{-20}$ &
        $1$&$.487\!\times\! 10^{-21}$ &
       $-3$&$.717\!\times\! 10^{-22}$ &
        $0$& \\
$49$ &  $1$&$.814\!\times\! 10^{-21}$ &
       $-5$&$.441\!\times\! 10^{-21}$ &
        $5$&$.441\!\times\! 10^{-21}$ &
       $-1$&$.814\!\times\! 10^{-21}$ &
        $9$&$.814\!\times\! 10^{-27}$ &
       $-1$&$.963\!\times\! 10^{-27}$ \\
$50$ &  $5$&$.735\!\times\! 10^{-22}$ &
       $-1$&$.211\!\times\! 10^{-21}$ &
        $7$&$.343\!\times\! 10^{-22}$ &
       $-1$&$.287\!\times\! 10^{-22}$ &
        $3$&$.217\!\times\! 10^{-23}$ &
        $0$& \\
\hline
\end{tabular}
\end{small}
\end{table}
\clearpage
\noindent
and so forth. Finally, this translates into an expansion for the
level-spacing distributions $P_{n}(s)$ via Eqs.~(\ref{eq:en}) and
(\ref{eq:pn}).

As Eq.~(\ref{eq:pn}) involves two derivatives with respect to the
variable $s$, we need to obtain the expansions of the functions
$\sigma_{n}(x)$ to two additional orders in $x$; so in order to
achieve a result for $P_{n}(s)$ that is correct to order $50$ in $s$
we have to compute the expansions of $\sigma_{n}(x)$ to order $52$ in
$x/\pi$. Furthermore, we need to expand the exponential function in
$D_{0}(s)$ (\ref{eq:d0exp}), whereas the integration on $x$ is
trivially performed on polynomials. As a result, the expansion
coefficients $p_{n;k}$ (\ref{eq:ps}) are expressions involving rather
lengthy rational numbers and powers of $\pi$. The expansions for
$P_{n}(s)$ with $0\le n\le 5$ are given in Appendix~\ref{app1}; for
higher $n>5$ the lowest order in the small-$s$ expansion of $P_{n}(s)$
is larger than $50$. The numerical values of the expansion
coefficients are given in Table~\ref{tab1}, compare also the numerical
values of the coefficients $p_{0;k}$ for $k\le 32$ given in
Ref.~\cite[Table~4.1]{Haake}.  The modulus of the coefficients
$p_{n;k}$ appears to decrease relatively rapidly with increasing order
$k$, so we may expect the series to converge on a rather large domain
around the origin.

\section{Numerical integration}

To give an account of the accuracy of the expansions, we compare them
to the result of a numerical integration of the coupled set of
differential equations (\ref{eq:pv0})--(\ref{eq:pv2}) derived from the
original Painlev\'{e} equation (\ref{eq:pv}). These are also obtained
using {\sc Mathematica} \cite{Wolfram}. In practice, we use initial
values for the functions $\sigma_{n}(x)$ for small, but non-zero $x$
which are obtained from their expansions, and then integrate the
system to larger values of $x$.

\begin{vchfigure}[b]
\includegraphics[width=0.5\textwidth]{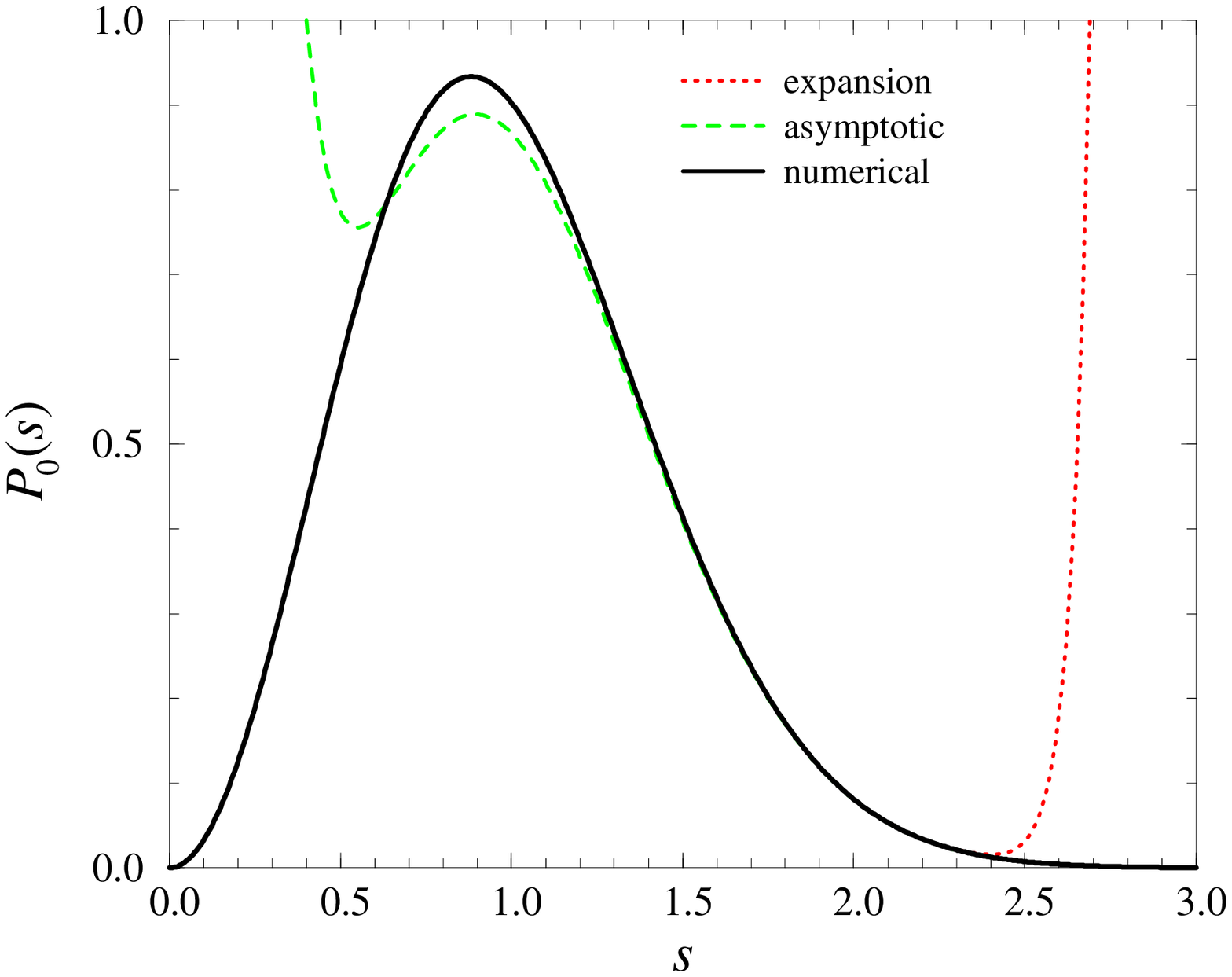}%
\vchcaption{The spacing distribution $P_{0}(s)$ as obtained
by numerical integration (solid line), the corresponding small-$s$ 
expansion (\ref{eq:pu0}) and the asymptotic form (\ref{eq:p0dys}).}
\label{fig:p0}
\end{vchfigure}

\begin{vchfigure}[t]
\includegraphics[width=0.5\textwidth]{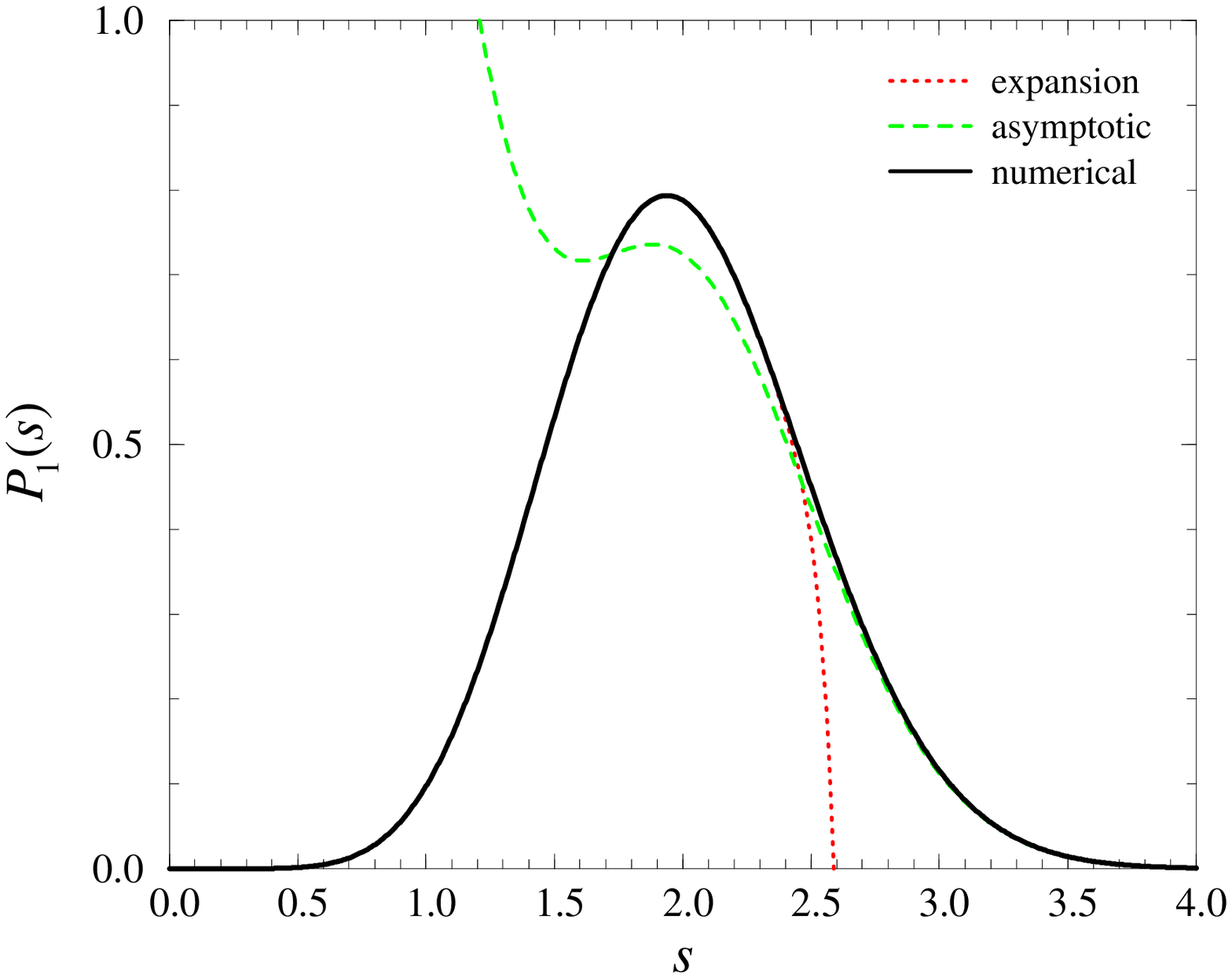}%
\vchcaption{The spacing distribution $P_{1}(s)$ as obtained
by numerical integration (solid line), the corresponding small-$s$ 
expansion (\ref{eq:pu1}) and the asymptotic form (\ref{eq:enbtw}).}
\label{fig:p1}
\end{vchfigure}

In fact, it turns out that this is not as easily done as stated. The
reason is that some of the functions involved become very small or
very large as $s$ is increased, and the computation of $P_{n}(s)$
essentially involves the cancellation of these large terms.  In
particular, the functions $\sigma_{n}(x)$ behave asymptotically like
\cite{BTW}
\begin{eqnarray}
\sigma_{0}(x)&\sim& -\frac{1}{4}x^{2}\\
\sigma_{n}(x)&\sim& -\frac{n!}{(8\pi)^{n/2}} \frac{\exp(nx)}{x^{n/2-1}}
\end{eqnarray}
for large argument $x$. So, for $n>0$, $\sigma_{n}(x)$ grows
exponentially with $x$, whereas $P_{n}(s)$ is at most of order one,
and eventually decreases as $\exp(-\pi^2 s^2/8)$ for large $s$, see
Appendix~\ref{app2} where the asymptotic behavior of the functions
$P_{n}(s)$ for large spacings $s$ is discussed.

\begin{vchfigure}[b]
\includegraphics[width=0.5\textwidth]{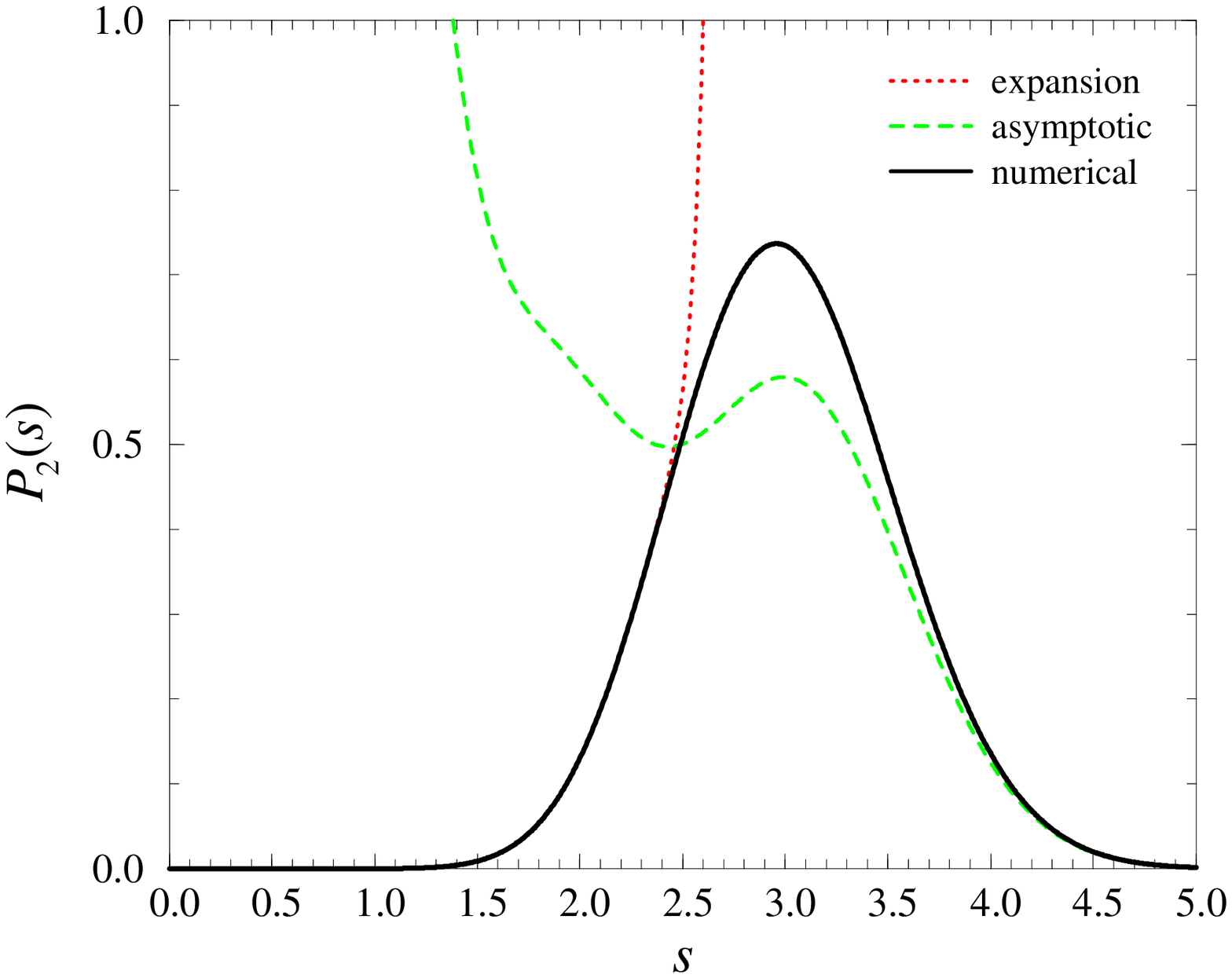}%
\vchcaption{The spacing distribution $P_{2}(s)$ as obtained
by numerical integration (solid line), the corresponding small-$s$ 
expansion (\ref{eq:pu2}) and the asymptotic form (\ref{eq:enbtw}).}
\label{fig:p2}
\end{vchfigure}

Results for $n=0$, $n=1$ and $n=2$ are shown in Figs.~\ref{fig:p0},
\ref{fig:p1} amd \ref{fig:p2}, respectively. Apparently, the numerical
solutions agree well with the series expansions of Appendix~\ref{app1}
up to about $s\lesssim 2.5$. For $n=0$, the distribution function is
well reproduced by the asymptotic form of Eq.~(\ref{eq:p0dys}) down to
$s\gtrsim 1.5$. Similarly, as shown in Figs.~\ref{fig:p1} and
\ref{fig:p2}, the small-$s$ expansions of Eqs.~\ref{eq:pu1} and
\ref{eq:pu2} describe the functions $P_{1}(s)$ and $P_{2}(s)$ for
$s\lesssim 2.5$. The asymptotic forms of
Eqs.~(\ref{eq:enbtw})--(\ref{eq:pnas}) reproduce the functions
$P_{1}(s)$ and $P_{2}(s)$ very well for $s\gtrsim 3$ and $s\gtrsim
4.5$, respectively. In principle, it should be possible to extend the
$s^{-1}$ expansion of Eq.~(\ref{eq:enbtw}) as well, at least for fixed
values of $n$, and thus improve the situation considerably.

\section{Concluding remarks}

We presented series expansions for the spacing distributions
$P_{n}(s)$ of the GUE. The expansion was calculated up to order $50$
in the spacing $s$, and for all distributions $P_{n}(s)$. In practice,
this means that for $0\le n\le 5$ the leading terms up to order
$s^{50}$ were obtained, whereas for $n>5$ the leading order is
larger. The expansions agree with the complete series solutions
obtained by Mehta \cite{Mehta92}, which were derived in a different
way.

The explicit expansions give a very precise account of the
small-spacing part of the spacing distributions, and might prove
useful for comparisons with experimental or numerical data. The
leading asymptotic behavior for large spacing was obtained in
Ref.~\cite{BTW} by similar means.

Within the same framework, albeit slightly more involved, one can
derive analogous series expansions for the spacing distributions of
the GOE and the GSE.

It is conceivable that other forms of expansions of the level spacing
distributions $P_{n}(s)$ for small $s$ might improve the convergence.
A candidate might be to start from a Wigner surmise type form as
discussed in Ref.~\cite{FW1}.

\begin{acknowledgements}
The author is indebted to C.~A.~Tracy who pointed out Ref.~\cite{BTW}
and provided him with the corresponding {\sc Mathematica}
programs. These established a starting point and inspired the
structure of the programs that were used to calculate the expansions
presented in this paper. The author also acknowledges useful
discussions with P.~J.~Forrester and R.~A.~R\"{o}mer. He also thanks
N.~S.~Witte for spotting some misprints in a couple of coefficients in
a previous version of this manuscript, and for bringing Refs.\
\cite{DiHa} and \cite{Mehta92} to his attention.
\end{acknowledgements}

\appendix 

\section{Leading terms of series expansions}
\label{app1}

Here we list the leading terms of the small-$s$ expansions for
$P_{n}(s)$.  The complete expansions can be downloaded as Mathematica
code from \cite{code}.  The numerical values of the coefficients up to
order $s^{50}$ are given in Table~\ref{tab1}.
\begin{eqnarray}
P_{0}(s) &=& \frac{\pi^{2}\,s^{2}}{3}
-\frac{2\,\pi^{4}\,s^{4}}{45} + 
\frac{\pi^{6}\,s^{6}}{315} - \frac{\pi^{6}\,s^{7}}{4050} -
\frac{2\,\pi^{8}\,s^{8}}{14175} + 
\frac{11\,\pi^{8}\,s^{9}}{496125} + 
\frac{2\,\pi^{10}\,s^{10}}{467775} -
\ldots
\label{eq:pu0}\\
P_{1}(s) &=&  \frac{\pi^{6}\,s^{7}}{4050} - 
\frac{11\,\pi^{8}\,s^{9}}{496125} + 
\frac{13\,\pi^{10}\,s^{11}}{13395375} -
\frac{4586\,\pi^{12}\,s^{13}}{170188239375} - 
\frac{\pi^{12}\,s^{14}}{2679075000} +
\ldots
\label{eq:pu1}\\
P_{2}(s) &=& \frac{\pi^{12}\,s^{14}}{5358150000} - 
\frac{17\,\pi^{14}\,s^{16}}{1181472075000} +
\frac{1577\,\pi^{16}\,s^{18}}{2859162421500000} - 
\ldots
\label{eq:pu2}\\
P_{3}(s) &=& \frac{\pi^{20}\,s^{23}}{9378525331350000000} - 
\frac{13\,\pi^{22}\,s^{25}}{1702202347640025000000} + \ldots
\label{eq:pu3}\\
P_{4}(s) &=& \frac{\pi^{30}\,s^{34}}{30645402510264863844600000000000} -
\ldots
\label{eq:pu4}\\
P_{5}(s) &=&
\frac{\pi^{42}\,s^{47}}{255963589608666174754500410100972300000000000000} -
\ldots
\label{eq:pu5}
\end{eqnarray}
For the level-spacing distributions $P_{n}(s)$ with $n>5$ the leading
terms are of higher order than $s^{50}$.

\section{Asymptotic behavior}
\label{app2}

The large-$s$ asymptotics of the level-spacing distributions $P_{n}(s)$
are also known. For $n=0$, Dyson's asymptotic result \cite{Dyson5,Haake}
reads
\begin{equation}
E^{\text{(a)}}_{0}(s) = 
\left(\frac{2}{\pi s}\right)^{1/4}
\exp\left(\frac{\ln 2}{12} + 3\zeta'(-1)-\frac{\pi^2 s^2}{8}\right)
\label{eq:e0dys}
\end{equation}
and thus
\begin{eqnarray}
P^{\text{(a)}}_{0}(s) 
&=& \frac{d^2}{ds^2}E^{\text{(a)}}_{0}(s)\nonumber\\
&=& \frac{\pi^4}{16}\left(s^2 - \frac{2}{\pi^2}
+ \frac{5}{\pi^4 s^2}\right) E^{\text{(a)}}_{0}(s).
\label{eq:p0dys}
\end{eqnarray}
Here, $\zeta'(-1)\approx -0.165421$ denotes the derivative of
Riemann's $\zeta$-function evaluated at $-1$. 

For $n>0$, the asymptotics are given in Ref.~\cite{BTW}. For the GUE, 
this gives
\begin{eqnarray}
\lefteqn{\frac{E^{\text{(a)}}_{n}(s)}{E^{\text{(a)}}_{0}(s)}
=  \frac{B_{n}\exp(n\pi s)}{s^{n^2/2}}\,
\Big(1 + \frac{(2n^{2}+7)n}{8\,\pi\, s} }
\nonumber\\
&& \hphantom{\times\Big(} 
+ \frac{(4n^{4} + 48n^{2} + 229)n^{2}}{128\,(\pi\, s)^{2}}
+ O[(\pi\, s)^{-3}]\Big)
\label{eq:enbtw}
\end{eqnarray}
where
\begin{equation}
B_{n} = 2^{-\frac{n^{2}+2n}{2}}\,\pi^{-\frac{n^{2}+n}{2}}\,
\prod_{m=1}^{n-1} m! \; .
\end{equation}
{}From this, we can easily calculate the corresponding asymptotic
behavior $P^{\text{(a)}}_{n}(s)$ as
\begin{equation}
P^{\text{(a)}}_{n}(s) \; = \; 
\frac{d^2}{ds^2} \; \sum_{m=0}^{n}\: (n-m+1)\: E^{\text{(a)}}_{m}(s),
\label{eq:pnas}
\end{equation}
which is just Eq.~(\ref{eq:pn}) applied to the asymptotic expressions.
The resulting terms are rather involved; however, it is apparent
from Eqs.~(\ref{eq:e0dys}), (\ref{eq:enbtw}), and (\ref{eq:pnas}) that
the leading behavior in all cases is $\exp(-\pi^{2}s^{2}/8)$.

\end{document}